**This is the accepted manuscript (postprint) of the following article:**
S. Mousavi, A. Moshfeghi, F. Davoodian, E. Salahinejad, *Eliminating the irregular surface layer of anodically-grown Ni-Ti-O nanopore arrays in a two-stage anodization*, Surface and Coatings Technology, 405 (2021) 126707.
https://doi.org/10.1016/j.surfcoat.2020.126707

# Eliminating the irregular surface layer of anodically-grown Ni-Ti-O nanopore arrays in a two-stage anodization

S.A. Mousavi, A. Moshfeghi, F. Davoodian, E. Salahinejad*

Faculty of Materials Science and Engineering, K. N. Toosi University of Technology, Tehran, Iran

**Abstract**

Nanopores (NPs) grown by anodizing can be partially hidden beneath a relatively compact surface oxide layer, which limits the volumetric surface area of these nanostructures. In this work, nitinol (NiTi) alloy was anodized in an aqueous electrolyte containing ethylene glycol, and sodium chloride in static and stirred electrolyte stages with the aim of removing the irregular surface array while achieving a thick NP layer. Electron micrographs showed that anodization in the static electrolyte provides a controlled thickness of NP layers covered by an irregular surface layer. In contrast, anodizing in the stirred electrolyte reduced the thickness and degree of irregularity, which were controlled by the different kinetics of dissolution at the tops, perimeters and bottoms of NPs. To benefit simultaneously from the thickness and regularity of the oxide layers, two-stage anodizing under static then stirred electrolyte conditions was found to be effective. Following a 30 min anodization in the static electrolyte, anodizing for 30 min under the stirred conditions provided the highest regularity in the oxide array, resulting in NPs of almost 40 nm and 11 μm in diameter and layer thickness, respectively. Two-stage anodizing under static then stirred electrolyte conditions is proposed in order to promote NP structures for applications demanding higher surface areas.

* Corresponding Author: Email Address: <salahinejad@kntu.ac.ir>

## 1. Introduction

The first successful fabrication of Ni-doped titania nanotubular structures (NTs) on the surface of NiTi alloy was performed by Kim et al. [1], the alloy being anodized in a fluoride-containing organic electrolyte. Due to their unique combination of physical properties, Ni-Ti-O NTs have shown a promising performance in electrocatalysis [2, 3], biosensing [4], gas sensing [5], and biomedical engineering [6, 7] applications. In many of these applications, an increase in the volumetric surface area of the NTs results in improved efficiency [8, 9]. Nevertheless, due to restrictions on the maximum obtainable thickness of Ni-Ti-O NTs in present electrolytes, a further increase in the volumetric surface area of the nanotubes is not possible. Recent studies have shown that the anodic fabrication of Ni-Ti-O NTs on NiTi alloy is morphologically limited to an approximately 1 μm thick layer in fluoride-containing electrolytes [10-12]. It has been concluded that this limitation is due to the rapid dissolution of nickel oxides in the electrolytes, leading to relatively short Ni-Ti-O NTs compared to pure anodically-grown $TiO_2$ nanostructures [13].

Nanopores (NPs) anodically grown on NiTi alloy are another form of high surface area nanostructures, alternative to NTs and not suffering from the aforementioned thickness limitation. In pursuit of obtaining Ni-Ti-O NPs with higher thickness and therefore higher volumetric surface area, numerous anodization experiments have been performed on NiTi with different compositions of organic-based electrolytes; all containing small amounts of $H_2O$ and differing in their etchant constituents: NaCl [14], HCl [15], NaBr [16], and $Na_2CO_3$ [17]. For example, in case of HCl-containing organic electrolytes, it has been shown that it is possible to produce NPs with a thickness of around 160 μm, successfully increasing the surface area of the nanostructure [15]. In addition to HCl, NaCl-containing electrolytes also promote the formation of thick NP layers on the NiTi substrate [14], while these nanostructures improve the corrosion resistance of the metallic substrate by limiting the access of corrosive media [11]. However, experiments on HCl- and NaCl-containing electrolytes have shown that the resulted NPs are partially hidden beneath a relatively compact surface oxide layer, limiting the surface area and efficiency of Ni-Ti-O NPs in their respective applications [18]. For example, the irregular surface layer formed during anodization of titanium substrates has shown to limit their efficiency in light absorption, a crucial property for dye-sensitized titania NT solar cells [19-21]. In case of Ni-Ti-O NP pseudocapacitors [1], the hindrance of electron and ion transfers by the irregular oxide layer adversely affects their performance. These observations encourage the pursuit of novel methods to eliminate this irregular oxide layer that particularly accompanies nanostructures on anodized NiTi alloy.

Different approaches have been explored for the removal of the irregular layer from the Ni-Ti-O NP tops produced in Cl-containing electrolytes. One of the most common strategies in fluoride-containing anodization experiments is to increase the anodizing time [22]. This provides further time for the self-organizing of the NP array to reach equilibrium and attain an improved morphological homogeneity. But in case of NiTi anodization in chloride-containing electrolytes, the increased duration of anodizing, even to 320 min, has proven to be futile in the elimination of the irregular surface oxide layer [18]. Other strategies include the use of a more viscous electrolyte base (glycerol) [16] and electrolyte stirring operation during anodization [23]. Using the more viscous glycerol instead of ethylene glycol (EG) commonly used, the irregular layer is successfully removed but the thickness of NPs is limited owing to the restricted mobility of ions in the highly viscous glycerol. Electrolyte stirring throughout anodizing in glycerol-based electrolytes has also proved to drastically shorten the final thickness of NPs. In the present work, for the first time, a simple, reproducible and low-cost process, including initial anodization in a static, chloride-containing electrolyte followed by anodization under electrolyte stirring, is introduced to fabricate thick NP layers free of the irregular surface layer on NiTi. The effect of different stirring times on the NP microstructure and morphology is imaged by electron microscopy and statistically analyzed.

## 2. Materials and methods

### *2.1. Materials*

NiTi alloy disks (Ni-50Ti, Kellogg's Research Labs, USA) of 9 mm diameter and 2 mm thickness were used as substrates. The electrolyte was prepared using ethylene glycol (Reag. USP grade, Merck, Germany), sodium chloride (>%99.5 NaCl, Merck, Germany), and double-distilled water (99% purity, 1 μS $cm^{-1}$ electrolytic conductivity).

### *2.2. Fabrication of NPs*

The NiTi disks were ground by abrasive papers, polished to a mirror finish and ultrasonicated in a bath containing ethanol and double-distilled water. The non-working side of the disks was connected to a copper wire then sealed to avoid any contact with the electrolyte. Each sample was anodized in 100 mL of an electrolyte containing ethylene glycol, 5 wt% double-distilled water and 0.3 M NaCl, based on a formulation in the literature [14]. A DC power supply was used to apply a cell voltage of 10 V between the working electrode and a Pt sheet cathode 20 mm away for controlled times at room temperature (25 ºC). To investigate the effect of electrolyte stirring on the microstructure and removal of the irregular surface layer, a combination of static and stirring-assisted anodization was used, where samples were initially anodized for 30 min in the static electrolyte. Anodization of the disks was continued in a stirred electrolyte for 10, 20 or 30 min, resulting in a total anodizing time of 40, 50 or 60 min, respectively. A cylindrical polytetrafluoroethylene-coated magnetic follower with a 4 mm diameter and 10 mm length centered in a 5 cm diameter cylindrical beaker containing 100 ml of the electrolyte and the rotational speed of 150 rev min$^{-1}$ were used to maintain controlled flow conditions in the electrolyte, while the 20 mm distanced electrodes were centered in the depth of 30 mm from the beaker bottom. To compare the samples, anodizing without stirring was also performed on the disks for a total time of 30, 40, 50 and 60 min. A single-step anodization under constant stirring for 30 min was also carried out.

*2.3. Characterization of NPs*

Field-emission scanning electron microscopy (FESEM, TESCAN, MIRA 3, accelerating voltage=:15 kV) after gold sputtering was used to assess the morphology of the specimens and to verify the removal of the irregular layer. Electron microscopy was carried on the anodized surfaces, the sub-surface and cross-sections of the samples. For the sub-surface observations, according to a procedure previously reported in the literature [13], a scratch was made on the as-anodized surface, using a steel cutter, to partially peel-off the NP layer and microscopy was carried out in the affected area. ImageJ software was also used to quantify the dimensions of the resulted nanostructures. Each statistical analysis on FESEM micrographs was performed on four distinct sections of the samples in terms of the mean and the standard deviation values of the diameter and number of pores and the thickness of the anodically-grown layer.

## 3. Results and discussion

Fig. 1 shows the surface, sub-surface and cross-sectional micrographs of the samples anodized in the static electrolyte at the cell voltage of 10 V. According to Fig. 1(a), the as-anodized surfaces are covered by an irregular layer, irrespective of the anodizing time. Fig. 1(b) shows a sub-surface micrograph of the as-anodized samples, revealing underlying regular NPs, i.e. NPs are partially hidden beneath the relatively compact surface layer. Although it has been shown that polishing of the substrate before anodization is effective in obtaining a uniform and aligned array of nanostructures [24], this work indicates the inadequacy of this treatment for nanostructures grown on NiTi (Fig. 1). The mean thickness of the NP layers formed on the anodized samples is also represented in the cross-sectional images of Fig. 1(c). By increasing the duration of anodizing, thicker NP layers were produced. Typically, 30 min of anodization results in NPs of approximately 17.7 µm thickness, whereas NPs with a maximum thickness of 24.5 µm are produced after 60 min, i.e.

only a 40% increase in thickness was achieved by doubling the anodization time. This suggests that as the thickness of NPs increased, their growth rate became slower, in agreement with previous studies [13]. This is likely to be caused by impeded mass transport of $Cl^-$ ions towards the electrolyte/oxide interface.

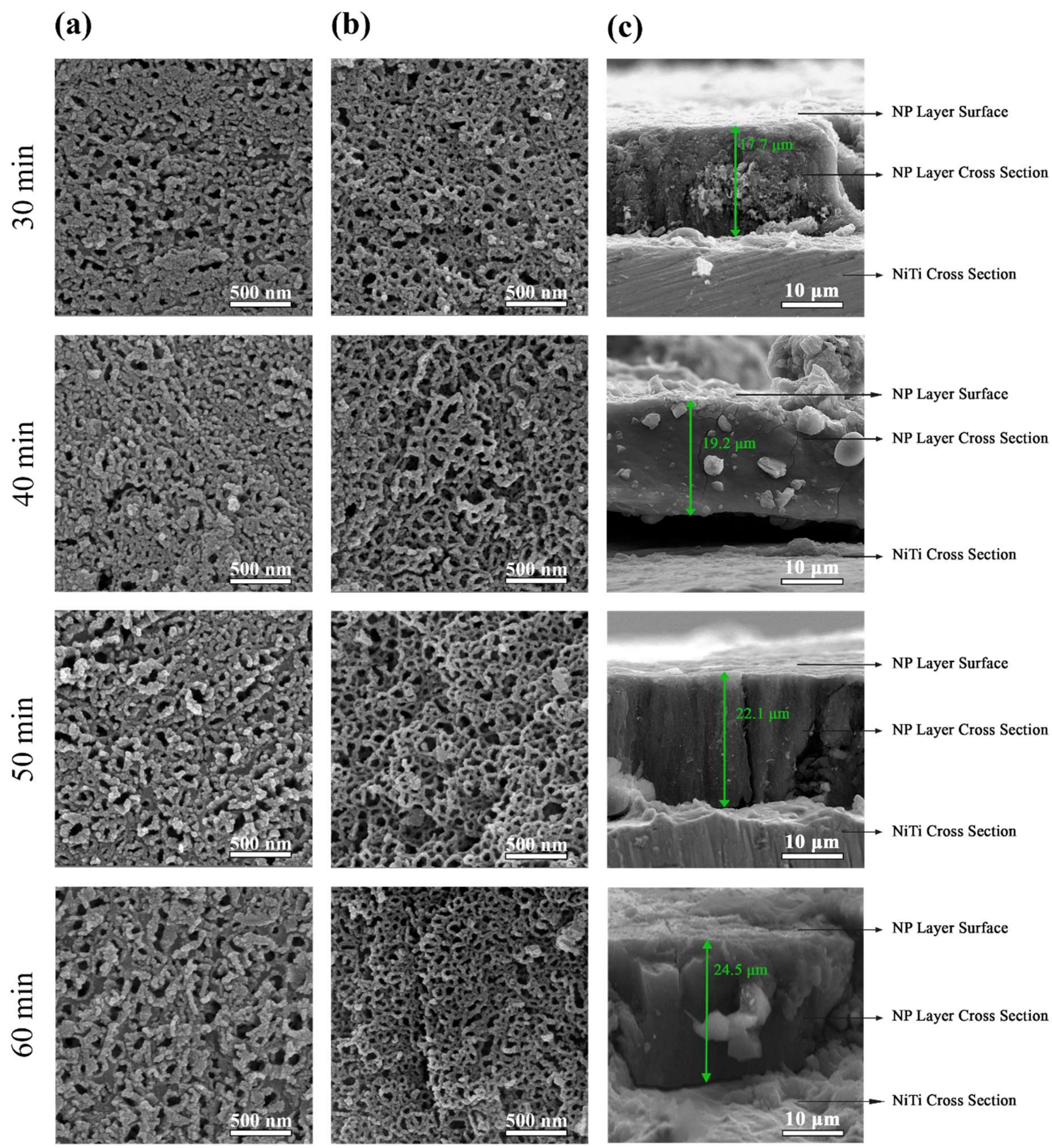

**Fig. 1.** (a) Surface, (b) sub-surface, and (c) cross-sectional FESEM images of the samples statically anodized for the different periods.

The formation of anodically-anodized NPs on the NiTi substrate is regarded as a self-assembly process governed by a balance between the oxide film formation and dissolution under actions of the etching ions. For Cl-containing electrolytes, the following reactions have been reported to be important during anodization [8, 11, 14, 16, 25]:

At the anode, titanium and nickel can form surface oxides:

$$Ti + 2H_2O \rightarrow TiO_2 + 4H^+ + 4e^- \quad (1)$$

$$Ni + H_2O \rightarrow NiO + 2H^+ + 2e^- \quad (2)$$

These oxides can be dissolved to form chlorocomplexes of titanium and nickel:

$$TiO_2 + 4H^+ + 6Cl^- \rightarrow [TiCl_6]^{2-} + 2H_2O \quad (3)$$

$$NiO + 2H^+ + 4Cl^- \rightarrow [NiCl_4]^{2-} + H_2O \quad (4)$$

It is also possible to form soluble metal complexes by metal cations reacting with excess chloride ions:

$$Ti^{4+} + 6Cl^- \rightarrow [TiCl_6]^{2-} \quad (5)$$

$$Ni^{2+} + 4Cl^- \rightarrow [NiCl_4]^{2-} \quad (6)$$

At the cathode, proton reduction results in hydrogen evolution:

$$2H^+ + 2e^- \rightarrow H_2 \quad (7)$$

As Reactions (1) and (2) continue, a relatively compact oxide layer can be formed at the metal/oxide interface, as shown in Fig. 2(a). The local anodic current density starts to drop as the oxide thickness increases. Simultaneously, $Cl^-$ ions can aggressively attack the oxide layer under the influence of electrical field, leading to formation of water-soluble $[TiCl_6]^{2-}$ and $[NiCl_4]^{2-}$ ions at the electrolyte/oxide interface [15, 16, 26] according to Equations (3) and (4). During anodic attack by $Cl^-$ ions, the compact oxide starts to breakdown at random locations, as shown in Fig. 2(b). With the subsequent rise of the current density, the breakdown locations act as nucleation sites for the growth of an irregular layer on the remaining compact layer, as indicated in Fig. 2(c). Later, according to Fig. 2(d), the high pH gradient between the top and bottom of the pores and the acceleration of Ti-Ni-O dissolution and pore penetration into the compact oxide layer occur at a lower pH [27]. This leads to the formation of NP arrays under the irregular surface layer. The field-assisted dissolution of the formed oxide layer can be characterized by three simultaneous reactions that take place at the NP tops, perimeters and bottoms, as schematically shown in Fig. 3. Dissolution at the NP tops of NPs leads to a reduction in the oxide thickness, whereas reactions at the perimeters and bottoms of NPs result in an increase in the NP diameter and thickness, respectively. The NP thickness in the samples anodized under the static electrolyte conditions can be rationalized by the dissolution of NP bottoms being dominant over the dissolution of their tops, since the NP thickness increases at a longer anodizing time. Subsequently, the growth rate becomes slower because it becomes more difficult for etching ions to reach the lengthened NP bottoms.

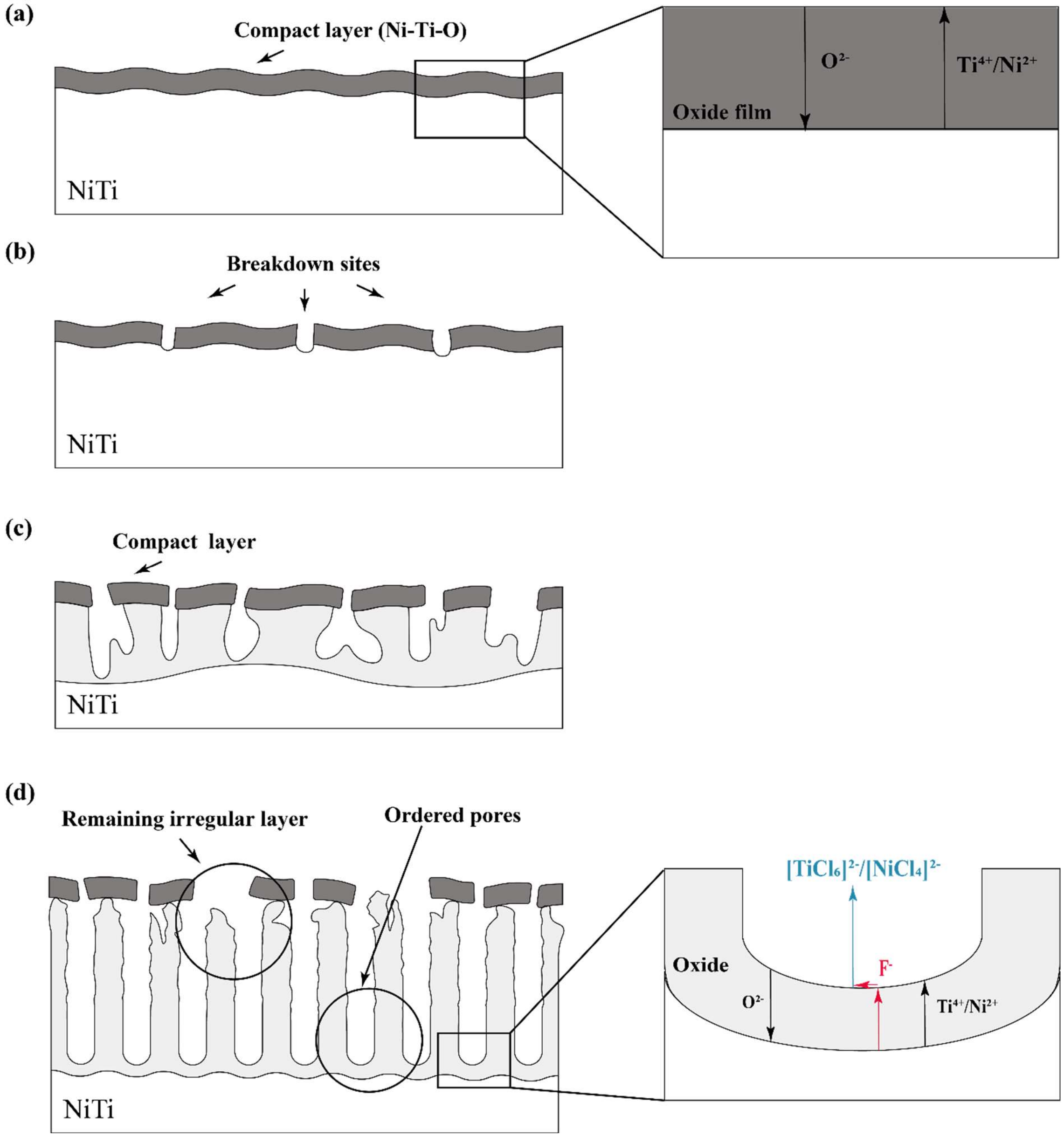

**Fig. 2**. Schematic illustration of the NP array formation: (a) formation of the compact oxide layer, (b) $Cl^-$ ion-assisted breakdown of the compact oxide layer, (c) following dissolution of the compact oxide layer which forms an irregular porous array and (d) formation of ordered nanostructures under the irregular surface by progression of anodization. Adapted from Ref. [21].

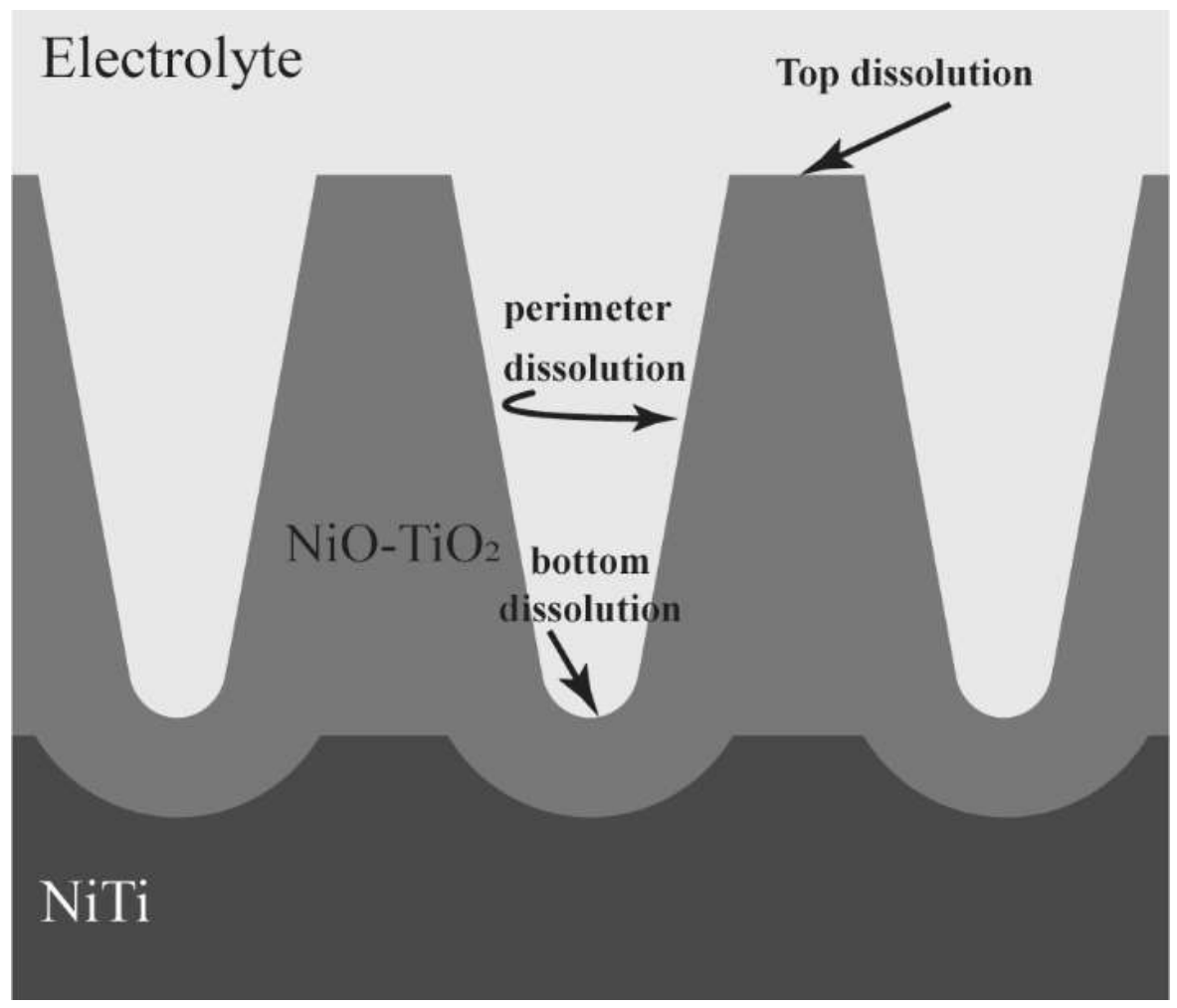

**Fig. 3.** Schematic illustration of the three dissolution types during anodization, where the electrolyte is supposed to only be in contact with the $NiO-TiO_2$ oxide layer.

Fig. 4 shows a micrograph of the samples anodized under the combined static, stirred electrolyte modes. According to Fig. 4(a) which illustrates the as-anodized surfaces, the irregularity of the formed NPs is visually lower than the surface micrographs of the samples anodized in the static electrolyte presented in Fig. 1 (a). The irregularity of structure is decreased as the duration of combined anodization is lengthened, so that joining of adjacent pores is observed in some areas. Fig. 4(b) is representative of the sub-surface micrographs, showing a greater visual similarity to the corresponding surface micrographs (Fig. 4(a)) compared to the similarity of Figs. 1(a) and 1(b). The NP thickness is also reported in Fig. 4(c), which clearly demonstrates that the thickness is shortened by additional stirring. More specifically, electrolyte stirring during anodizing leads to the following trends in the observed NP thickness:

*i*) Following anodizing in a static electrolyte for 30 min, the NP thickness is progressively shortened when the duration of the electrolyte stirring is increased from 0 to 30 min.

*ii*) The sample anodized in the stirred electrode for 30 min shows a lower thickness of NPs compared to the sample anodized in a static electrolyte for 30 min.

*iii*) The thickness of NPs in the samples anodized under the static and stirring conditions is lower than their counterparts in static electrolytes, i.e. the combined 40 (30 min of static + 10 min of stirring), 50 (30 min of static + 20 min of stirring) and 60 (30 min of static + 30 min of stirring) min relative to purely static 40, 50 and 60 min processes, respectively.

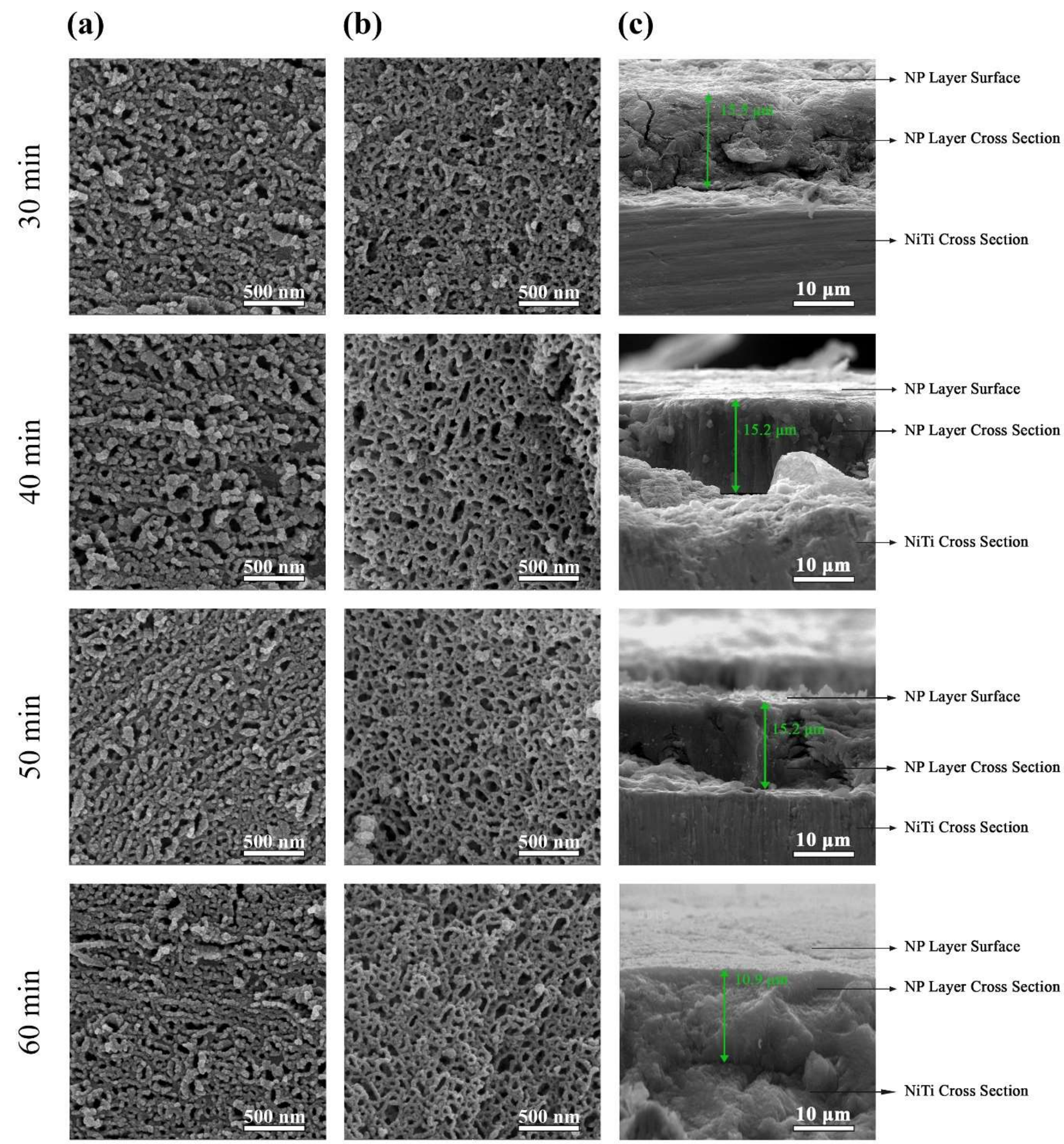

**Fig. 4.** (a) Surface, (b) sub-surface, and (c) cross-sectional FESEM images of the samples processed in the purely stirred electrolyte for 30 min and by the combined anodization modes of 40, 50 and 60 min.

The formation and growth of NPs are charge transfer controlled at the NiTi surface, or mass transfer controlled by diffusion of ionic species towards the anode surface. The decrease in the growth rate of the NP layer at longer anodization in the static electrolyte (Fig. 1) is indicative of the difficulty of $Cl^-$ ion diffusion to the NP bottoms as these anions are a major contributor to NPs growth. This can be interpreted in terms of a mass transfer controlled growth of NPs or concentration polarization in front of the anode, as evidenced by the current density drop in similar studies [14]. The chemical dissolution of oxides at the NP tops is believed to be a reason for the decrease in the NP layer thickening rate [23, 28], which is not affected by concentration polarization to the same extent as dissolution at the NP bottom due to the easier access to the electrolyte. In the absence of stirring and before the onset of the thickening of the diffusion layer, the chloride ions tend to dissolve the oxide at the NP bottoms due to the stronger electric field. This confirms the assignment of the ion transfer controlling the reactions rate under the static conditions. Under electrolyte stirring, the domination of diffusion and the polarization concentration are diminished and the rate of dissolution at the NP bottoms and tops increases. Electrolyte stirring can contribute to a reduction in the local temperature of the electrolyte near the anode, resulting in an increase in the electrolyte viscosity in comparison to the static state and in a lower rate of mass transport [29]. In combination with the lower field strength at the electrolyte/oxide interface due to the reduced thickness of the NP layer after the prior anodizing in the static electrolyte, it results in the predomination of NP top dissolution over NP bottom dissolution and a more decrease of the NP layer thickness on stirring. In contrast, stirring during anodization of other valve metals has usually led to a higher thickness for corresponding oxide nanostructures [30-33], due to the importance of diffusion in steady-state growth of these nanostructures during anodizing in viscous electrolytes (e.g. ethylene glycol or glycerol) containing corrosive ions [34, 35]. For samples subjected to combined static, stirred electrolyte anodizing stages, a slower NP bottom dissolution during the subsequent electrolyte stirring is attributed to further restricted ion transfer through NPs formed in the prior static anodization process. This is a reason for the higher thickness of NPs formed on the 30 min stirred sample in comparison to the combined 60 min sample for which the subsequent stirring seems to be mainly involved with the NP top dissolution rather than the NP bottom dissolution.

Fig. 5 represents the diameter of NPs anodically grown under the different conditions, extracted statistically from the surface micrographs (Figs. 1 and 4). As can be seen, the increase of the anodization time in the static electrolyte slightly increases the mean diameter of NPs, albeit with no significant difference. This reflects the minor dominance of the NP perimeter dissolution over the NP top dissolution, due to the fact that the effect of field-assisted dissolution diminishes in intensity as the distance between the positively-charged substrate and the etching ions is increased. Additionally, it is believed that some NPs join together by progression of anodization, generating larger NPs on some areas of the surface. Regarding the samples subjected to the additional stirring operations, the mean diameter of NPs initially increases in the first 10 min of stirring, but a following decrease is evidenced again with an insignificant difference. Upon the initiation of electrolyte stirring, the NP top dissolution is believed to be drastically increased and to prevail over the level of the perimeter dissolution. Accordingly, due to the conical nature of anodically-grown NPs [8] as illustrated in Fig. 3, the mean diameter of NPs is overall decreased by increasing the electrolyte stirring time. It can be also seen that the standard deviation of the NP diameter for all the samples is at high levels, which is due to the scattered distribution of the diameters. In this regard, Fig. 6 depicts the NP diameter distribution curve of the selected samples (60 min static and 60 min combined conditions). In agreement with the standard deviations, the string operation involved in the anodization process provides a narrower diameter distribution, which is indicative of becoming regular. This conclusion is also verified by the qualitative observations on the micrographs, as described for Figs. 1 and 4.

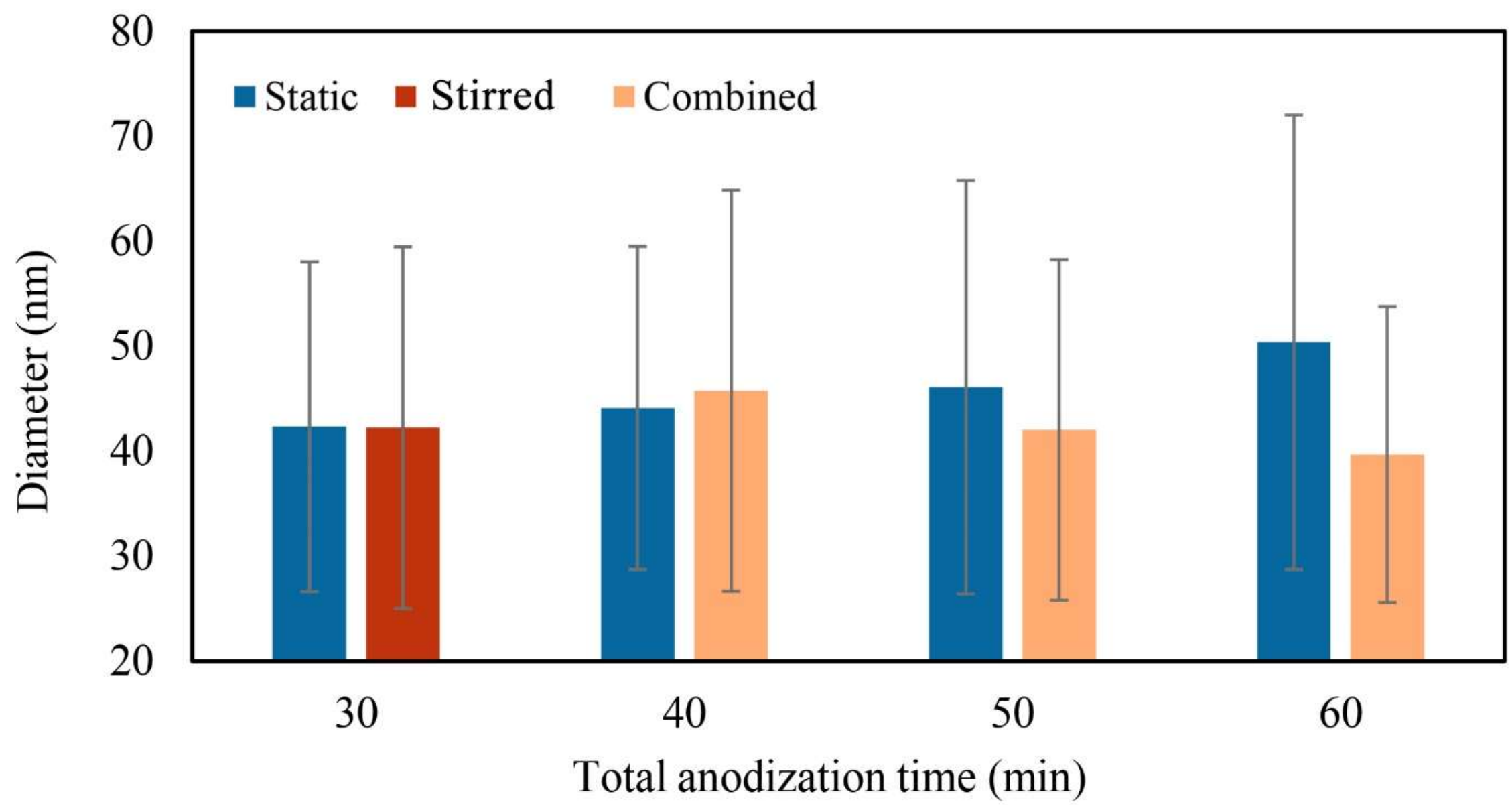

**Fig. 5.** Diameter of NPs anodically fabricated under different stirring conditions.

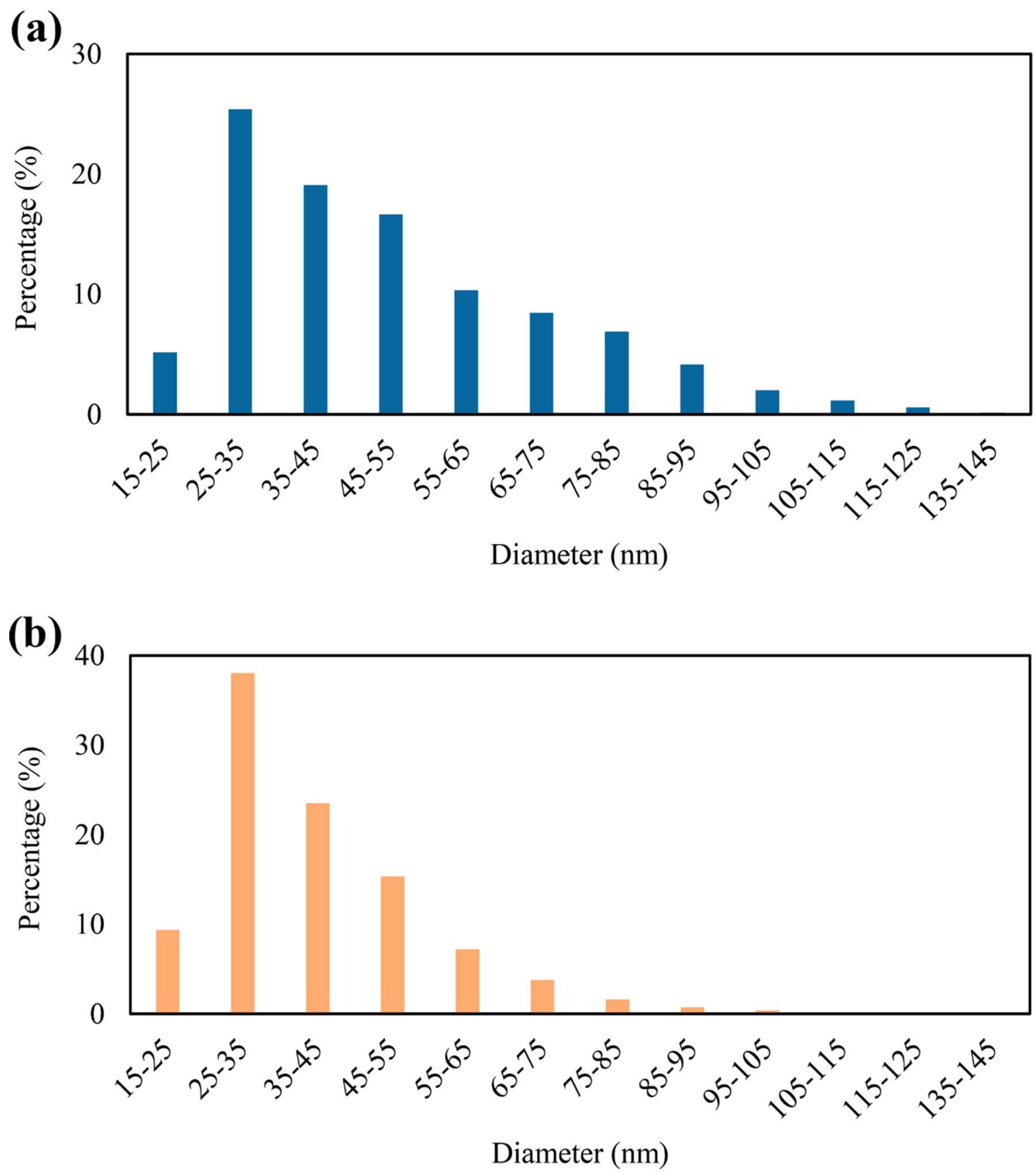

**Fig. 6.** Histogram of the NP diameter distribution for (a) the 60 min statically anodized sample and (b) the sample anodized in the static electrolyte for 30 min and the subsequent stirred electrolyte for 30 min.

Fig. 7 is illustrative of the total number of NPs over a surface area of 2 ×2 $\mu m^2$ for the differently anodized samples. As another measure of the irregularity level of the anodized surfaces, the standard deviation value for the number of NPs is also considered. For the samples anodized in the static electrolyte, the mean total number of NPs is almost constant up to 50 min of anodization and thereafter, a sudden decrease is observed. This instantaneous reduction in the mean total number of NPs can be ascribed to the increase of the NP perimeter dissolution and/or the joining of some adjacent pores accompanied by the generation of larger pores. This assumption is further corroborated by the increase detected in the mean diameter of NPs from 50 min to 60 min of anodization (Fig. 5). As realized from the standard deviation of the NP number, the irregularity level of the surface layer is also perceived to decrease in the first 50 min of static anodization, followed by increase due to the generation of larger NPs. For the samples anodized under the combined static, stirred electrolyte conditions, the mean total number of NPs is increased with the total anodization time, while the increase from 50 min to 60 min is not statistically significant. The abrupt increase can be attributed to a decrease in the irregularity of the surface layer. From the perspective of the standard deviations for the combined conditions, it is seen that the highest value (49.7) as the most heterogeneous distribution of NPs is related to 50 min of anodization, which is attributed to the partial removal of the irregular layer with respect to the total sample surface. By increasing the duration of the stirred-electrolyte anodization (60 min), it is observed that the standard deviation decreases to 3.9 as the irregular layer is mostly removed from the sample surface.

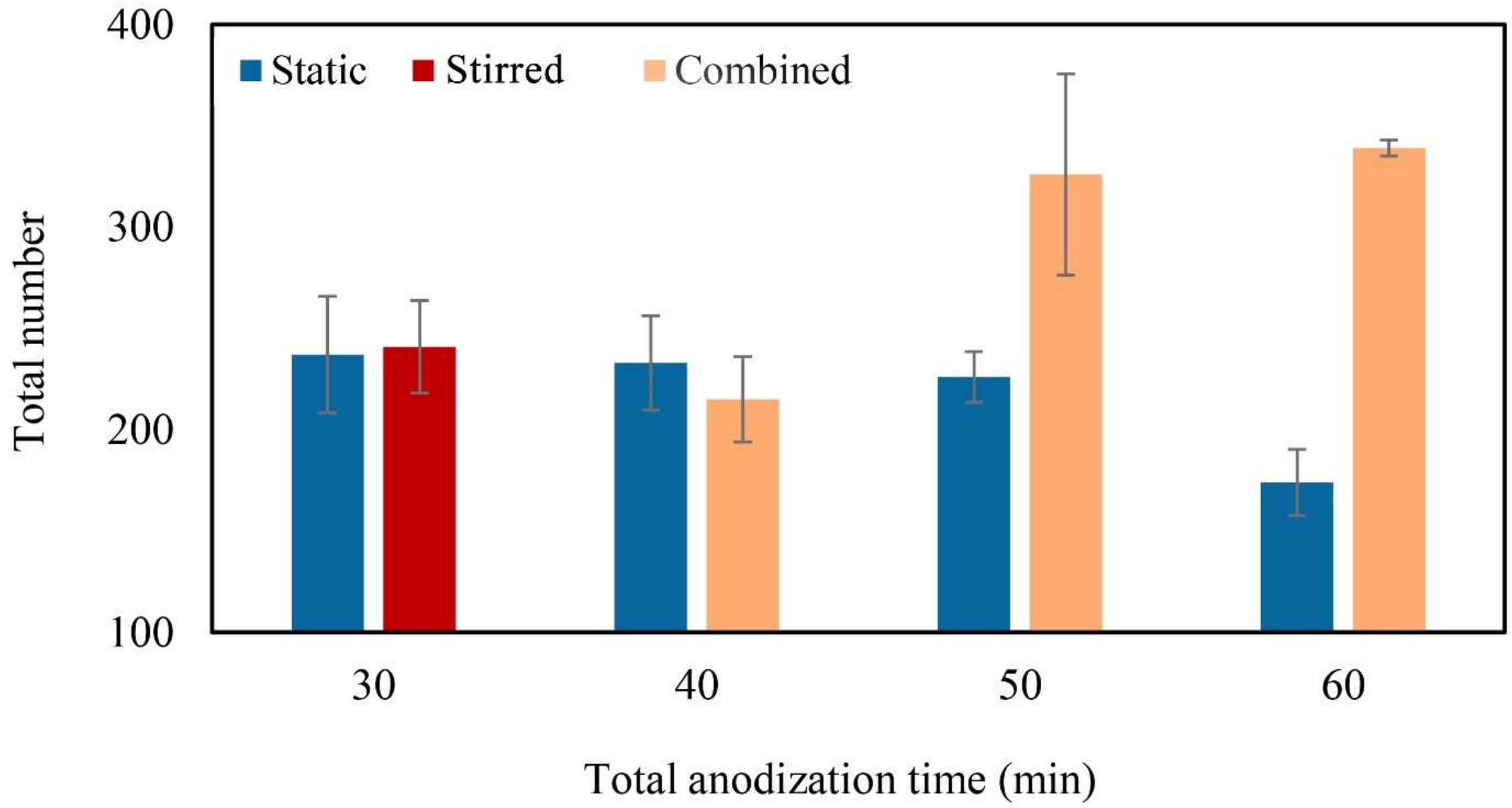

**Fig. 7.** Total number of NPs formed on the different samples over the surface area of 2×2 μm$^2$.

The sub-surface micrographs of the selected samples were also statistically analyzed. To quantify the regularity level, a parameter, *R* is used:

$$R = \frac{\text{Mean number of NPs in surface}}{\text{Mean number of NPs in sub-surface}} \quad (8)$$

Clearly, the convergence of this fraction to unit is indicative of similarity between the surfaces and sub-surfaces due to the removal of the irregular layer. The *R* values calculated for the 30 min static, 50 min static and 60 min combined conditions are 0.70, 0,72 and 0.97, respectively, which confirms the comparisons drawn above on the regularity. Typically, the 60 min combined static, stirred electrolyte sample is considered to achieve the best result regarding the NP array formation with a blocked area as small as 3%. The mean diameter of NPs in the sub-surface of the 30 min statically, 50 min statically and 60 min static then stirred electrolyte anodized samples is estimated to be 42, 44 and 39 nm, respectively. In comparison, the lower diameter of NPs detected in the sub-surfaces than the surface counterparts is another evidence for the conical geometry of anodically produced NPs on NiTi (Fig. 3), similar to NPs produced on titanium alloys [8, 36].

## 4. Conclusions

This work focuses on the irregular surface layer formed on anodically anodized nitinol which undesirably blocks some underlying NPs. It was found that the irregular surface layer was not removed by increasing the anodization time in a static electrolyte. However, an additional electrolyte stirring stage after 30 min of static anodization could achieve homogenous NP structures. 30 min anodization in the stirred electrolyte mainly eliminated the irregular surface layer and yielded a nanoporous layer of 11 μm in thickness, which seems to be promising for relevant applications demanding a high volumetric surface area.